\documentclass[aip,showpacs]{revtex4}
\usepackage{graphicx}

\begin{document}
\def \e{\epsilon}

\title{A non-trivial PT-symmetric continuum Hamiltonian and its Eigenstates
and Eigenvalues} 



\author{Lawrence R. Mead}
\email{lawrence.mead@usm.edu}
\affiliation{School of Mathematics and 
Natural Sciences, The University of Southern Mississippi, Hattiesburg MS 39406-5045, USA}
\author{David Garfinkle}
\email{garfinkl@oakland.edu}
\affiliation{Dept. of Physics, Oakland University Rochester, MI 48309}
\author{Sungwook Lee}
\email{sunglee@usm.edu}
\affiliation{School of Mathematics and 
Natural Sciences, The University of Southern Mississippi, Hattiesburg MS 39406-5045, USA}


\date{\today}

\begin{abstract}
In this paper, a non-trivial system governed by a continuum PT-symmetric
Hamiltonian is discussed. We show that this Hamiltonian is iso-spectral to
the simple harmonic oscillator. We find its eigenfunctions and the path
in the complex plane along which these functions form an orthonormal set. We
also find the hidden symmetry operator, ${\cal C}$, for this system. All 
calculations are performed analytically and without approximation.
\end{abstract}


\maketitle 

\section{Introduction}
Over the last two decades there has been considerable interest in PT-symmetric
systems after Carl M. Bender and collaborators discovered that a quantum
Hamiltonian need not be Hermitian to have real eigenvalues, but could also
be PT-symmtric.\cite{1} Here, {\cal P} stands for parity, {\cal T} for time
reversal (complex conjugation). It was found that there is a hidden symmetry
operator for such systems which does not exist in Hermitian systems. The
corresponding operator, {\cal C}, is necessary in order to define a new inner 
product for which the eigenstates of the Hamiltonian have positive norm. 

Since this discovery, PT-symmetric quantum theory has entered the main stream
of physics investigations and the number of articles written by many authors 
on PT-symmetric systems is legion.\cite{2}
The consistency of PT-symmetric quantum
mechanics has been shown; \cite{3} PT-symmetry has appeared in quantum optics
\cite{4,5} as well as technology \cite{6,7,8}. Recently, even time-dependent
systems have been studied.\cite{9} An excellent text on PT-symmetry
by Bender was published in 2019 (with many contributors and dozens of
references) which will be our main reference for this article.\cite{10}

Here, we will present a non-trivial system governed by a PT-symmetric
Hamiltonian in the continuum. In contrast to non-trivial systems studied
in the past, such as the $ix^3$ oscillator, for which perturbation theory
has been required, we will perform all calculations without
approximation. We have organized the paper as follows. In the next section,
we will review the relevant and important facts on PT-symmetric quantum
theory. The Hamiltonian will then be presented. Next, we present a mathematical theorem which will allow us to further analyze this Hamiltonian and show that 
it is isospectral to the simple harmonic oscillator. In a further section, we 
will derive the exact eigenstates of the Hamiltonian and demonstrate the path 
in the complex plane along which one must integrate to ensure the states form 
an orthonormal set. Finally, we will derive the hidden symmetry operator for 
this system.
\section{Brief summary of PT-symmetric quantum theory}
A PT-symmetric Hamiltonian satisfies
\begin{equation}
[H,{\cal PT}]=0,
\end{equation}
where P is parity and T is time reversal. The latter means complex conjugation
in order to preserve the fundamental commutation relation $[x,p]=i$ under
PT (we take $\hbar=1$ throughout): under PT, $p \to p$ and $x \to -x$. 
Examples of (time-independent) potentials might be $V(x)=ix^3$ or 
$V(x)=x^2(ix)^n$. Eigenstates of such a Hamiltonian will be denoted here as 
$\phi_n(x)$; thus,
\begin{equation}
H\phi_n(x)=E_n \phi_n(x).
\end{equation}
Such states do not have norms which are preserved in time. The so-called
PT-norms are preserved,
$$N=\int \phi_n(x) {\cal PT} \phi_n(x) dx,$$
but they can be {\it negative} which is not allowed for a probability. A
norm may however be defined as follows. There is a hidden symmetry which
is expressed by a new operator, $\cal C$, which does not exist in Hermitian
quantum theory. The action of this new operator is
\begin{equation}
{\cal C} \phi_n(x) = (-1)^n \phi_n(x),
\end{equation}
corresponding to the PT-norm  $\pm 1$. Thus, the CPT-norm,
\begin{equation}
N=\int \phi_n(x) {\cal CPT} \phi_n(x)
\end{equation} 
is real and {\it positive} for all $n$. The method of finding this operator
is typically to note that it must satisfy
\begin{equation}
 [H,{\cal C}]=0=[{\cal C},{\cal PT}] \qquad {\cal C}^2 = 1.
\end{equation}
These equations have been solved perturbatively or by semi-classical means
in some non-trivial cases.\cite{11,12} It is known that the C-operator may
be written as,
\begin{equation}
{\cal C}=e^{Q(x,p)}{\cal P},
\end{equation}
where the operator $Q(x,p)$ is {\it even} in position, $x$, and {\it odd}
in the momentum, $p$, in order to satisfy the above requirements. We will
freely use this representation in what follows. Also, it has been pointed 
out that for a PT-symmetric Hamiltonian, there exists an iso-spectral
Hermitian Hamiltonian, $h$ with corresponding eigenstates $\psi_n(x)$ both
related to their PT counterparts by a similar exponential transformation:
\begin{equation}
h=e^{-Q/2} H e^{Q/2}, \qquad \psi_n(x)=e^{-Q/2} \phi_n(x).
\end{equation}
The Schr\"odinger equation
$$h\psi_n(x)=E_n \psi_n(x)$$ is easily shown to be identical to that in the
PT-symmetric realm, (2), by direct substitution; the eigenvalues, $E_n$ are the
same in both. Eigenstates of $h$ are orthonormal in the usual way, 
as are the eigenstates of $H$, but in the PT-symmetric realm, it is with 
respect to the CPT-inner product:
\begin{equation}
\int \phi_n(x) {\cal CPT} \phi_m(x) dx = \delta_{nm}.
\end{equation}
Finally, one must note that this latter integral need not be evaluated along
the real axis, but may need to be along a curve in the complex x-plane. We
will address this need in our example.

\section{The Hamiltonian}
The PT-symmetric Hamiltonian we will discuss is
\begin{equation}
H={\frac 1 2}p{s^4}p + 4\e^2{s^2} +{\frac {x^2} {2{s^2}}},
\end{equation}
where $s=1+2i\e x$, $\e$ is a real parameter and as usual, $p=-id/dx$ is
the momentum operator. We will use the notation $s$ for $1+2i\e x$ throughout
as well as its complex conjugate $\bar s=1-2i\e x$. That this Hamiltonian
is PT-symmetric is manifest because $PT s = s$; and it
will reduce to the simple harmonic oscillator at $\e=0$. In fact, this
PT-symmetric Hamiltonian is {iso-spectral} to the harmonic oscillator with
eigenvalues $n+1/2$ (for convenience, we will take the oscillator frequency
to be $\omega=1$ and particle mass $m=1$.). In order to demonstrate this
last assertion, we will need a mathematical lemma which we now state (this
lemma will be used several times in what follows):\\
\medskip
{\bf LEMMA} - For any differentiable function $U(x)$, real parameter $\e$
and operator $F=x^2 p + p x^2$,
\begin{equation}
e^{\e F} U(x) = {1\over s} U(x/s).
\end{equation}
In order to prove this lemma, we begin with the expansion of the exponential.
\begin{equation}
e^{\e F}U(x)=\sum_{n=0}^\infty {\e^n\over n!}F^{(n)}U(x)\equiv
\sum_{n=0}^\infty f_n(x).
\end{equation}
Before continuing, we note that $f_0(x)=U(x)$. Now, the $n+1$th term in this
series satisfies
$$f_{n+1}(x)={\e \over (n+1)}F f_n(x),$$ or, after substituting the form 
of the operator $F$,
\begin{equation}
f_{n+1}(x)=-{\mu \over n+1} [xf_n(x)+x^2 f'_n(x) ].
\end{equation}
with $\mu = 2i\e$
We note that this is a differential-difference equation. To solve this we 
multiply this result by $z^{-n}$ and sum over all $n$ from $0$ to $\infty$ and
define
\begin{equation}
g(x,z)=\sum_{n=0}^\infty f_n(x) z^{-n}.
\end{equation}
$g(x,z)$ constitutes the z-transform of $f_n(x)$ which is invertable in
terms of a contour integral. We will not need the explicit form of 
$f_n$, but only the properties listed below of $g(x,z)$ itself:
\begin{equation}
\lim_{z\to \infty}g(x,z)=f_0(x)=U(x), \qquad e^{\e F}U(x)=g(x,1).
\end{equation}
Now, according to Eq.(12,13), $g(x,z)$ satisfies the first-order PDE
\begin{equation}
z^2{\partial g(x,z)\over\partial z}-\mu x^2 {\partial g(x,z) \over \partial x}
=\mu x g(x,z).
\end{equation}
This PDE is solvable by standard means - the exact, general solution is
\begin{equation}
g(x,z)={1\over x}\Phi({2 i\e x z\over z+2i\e x}).
\end{equation}
We must satisfy the initial condition $g(x,\infty)=U(x)$; hence
\begin{equation}
U(x)=\lim_{z\to \infty} g(x,z)={1\over x}\Phi(2i\e x)\equiv {1\over x} \Phi(A),
\end{equation} and thus
$$\Phi(A)={A\over 2i\e}U({A\over 2i\e }),$$ or
$$g(x,z)={z\over z+2i\e x}U({xz \over z+2i\e x}).$$ The series sum is
$g(x,1)$; hence the lemma follows
\begin{equation}
e^{\e F} U(x)=g(x,1)={1\over s} U(x/s)
\end{equation} 
\\
The simple harmonic oscillator Hamiltonian is 
$h={1\over 2}p^2 +{1\over 2}x^2.$
With this lemma in hand, we may now show that whenever $\psi_n$ is a solution
of the harmonic oscillator $h\psi_n=E_n \psi_n$ then the corresponding
PT-symmetric $H$ above satisfies $H\phi_n=E_n\phi_n$ with the same $E_n$.
To simplify the coming algebra, let $w=1/s$, $z=wx$ and $s$ is defined
as before. Now, since the action of $p$ on $x$ is $-i$ it follows that
\begin{equation}
p w = -2\e w^2, \qquad p z = -iw^2
\end{equation}
The Hamiltonian (9) in this notation is
\begin{equation}
H={1\over 2} p w^{-4}p+4 \e^2 w^{-2}+{1\over 2}z^2.
\end{equation}
Let us suppose that the transformation between $\phi$ and $\psi$ is given
by Eq.(7) and the lemma as
\begin{equation}
\phi(x)=e^{Q/2} \psi(x) = e^{\e F} \psi(x) = w\psi(z),
\end{equation}
in the current notation and
where the transformation operator is $Q=2\e F$ (later, we will show this
explicitly in another way). We find using the above that
\begin{equation}
p\phi = \psi p w + w{d\psi \over dz} p z = -iw^3 {d\psi\over dz}-2\e w^2 \psi.
\end{equation}
We then find, using straightforward algebra that
\begin{equation}
{1\over 2}pw^{-4}p \phi=w \Big [ -{1\over 2} {d^2\psi \over dz^2}-
4\e^2 w^{-2}\psi \Big ].
\end{equation}
Then it follows that
\begin{equation}
H\phi = w \Big [-{1\over 2} {d^2 \psi \over dz^2} + {1\over 2}z^2 \psi(z)
\Big ]
\end{equation}
The expression in square brackets is just the usual harmonic oscillator
Hamiltonian acting on $\psi$. Thus, if $\psi$ is an eigenstate of the 
oscillator, then $\phi=\exp(\e F) \psi$ is an eigenstate of the PT-symmetric
Hamiltonian with the same eigenvalue. We will later write down the states
$\phi_n(x)$ explicitly.

The Hamiltonian (9) may be obtained from the harmonic oscillator directly.
Reversing the transformation given in (7) we have
\begin{equation}
H=e^{Q/2} h e^{-Q/2}=e^{Q/2} [ {1\over 2}p^2 + {1\over 2} x^2 ] e^{-Q/2}
\end{equation}
The right-hand side may be written in terms of nested commutators with
$Q=2\e F=2 \e [x^2 p + p x^2]$:
\begin{equation}
H=e^{\e F} h e^{-\e F}= h +{\e \over 1!}[F,h]+{\e^2 \over 2!}[F,[F,h]]+\dots
\end{equation}
or
\begin{equation}
H=h+{\e \over 1!}C_1+{\e^2 \over 2!} C_2 + \dots,
\end{equation}
where $C_n=[F,C_{n-1}]$.
As one calculates commutators, one notices two types of terms: powers of $x$,
or powers of $x$ sandwiched between two factors of momentum, $p$.
\begin{equation}
[F,x^n]=-2inx^{n+1}, \quad [F,px^np]=i(8-2n)px^{n+1}p-2nix^{n-1}
\end{equation}
It turns out that there are only a finite number of terms bounded by $p$s,
a finite grouping of powers of $x$ and an infinite series in powers of
$x$ alone. This series may be easily summed, and combining all terms yields
the Hamiltonian (9).
\section{The eigenfunctions of $H$}

The eigenfunctions of $H$ may be found directly from the transformation (7).
Recall that the wavefuntions of the simple harmonic oscillator are
\begin{equation}
\psi_n(x)=A_n H_n(x) e^{-x^2 /2},
\end{equation}
where the $A_n=(\sqrt{\pi}2^n n!)^{-1/2}$, $H_n(x)$ are Hermite polynomials
satisfying $H_n(-x)=(-1)^n H_n(x).$
Thus, by our lemma
\begin{equation}
\phi_n(x) = e^{\e F}\psi_n(x)={A_n \over s} H_n(x/s) e^{-x^2/2s^2}
\end{equation}
We note that due to the symmetry property of the $H_n$ noted above 
${\cal PT}\phi_n(x)=(-1)^n \phi_n(x)$; that is, these are eigenstates of
PT with eigenvalues $\pm 1$.

We may now write down the C-operator for this system and find its action on
these states. According to Eq. (6), the C-operator should be
\begin{equation}
{\cal C} = e^Q {\cal P} = e^{2\e F}\cal P.
\end{equation} 
We have already shown the action of the exponential operator above in the lemma;
we need only double the coefficient of $F$. Hence,
\begin{equation}
{\cal C}\phi_n(x)=e^{2\e F} \phi(-x)={1\over t}\phi_n(-x/t),
\end{equation}
where $t=1+4i\e x$ (similar to $s$). In order to find the explicit action
of $C$, we need only calculate the following:
\begin{equation}
{\cal C} [{1\over s}]=[{1\over t}][{1\over 1-2i\e x/t}] = {1\over s},
\end{equation}
\begin{equation}
{\cal C} \Big [ {x\over s}\Big ]^n=\Big [{-x/t \over 1-2i\e x/t} \Big ]^n =
(-1)^n [x/s]^n.
\end{equation}
It then follows immediately from the explicit form of $\phi_n(x)$ above
that
\begin{equation}
{\cal C} \phi_n(x)=(-1)^n \phi_n(x),
\end{equation}
as it should according to the general theory. The action of $\cal{CPT}$
is thus to  multiply the state $\phi_n$ by unity and thus the CPT-norm of
the state will be positive.

The last property we need to demonstrate is the orthogonality of the states
in (30). We begin with the known orthonormality of the harmonic oscillator
states (which are real-valued):
\begin{equation}
\delta_{nm}=\int_{-\infty}^\infty \psi_n(q) \psi_m(q) dq.
\end{equation}
Transforming these states to eigenstates of $H$ by Eq.(7) using our lemma,
we have
\begin{equation}
\delta_{nm}=\int_{-\infty}^\infty {dq \over \bar s^2}\phi_n(q/\bar s)
\phi_m(q/\bar s),
\end{equation}
with $\bar s=1-2i\e q$.
In order to make sense of this orthogonality relation, we change variables
from the real number, $q$, which runs over all real values, to the complex 
number $z$ which is parametrized by $q$:
\begin{equation}
z=x+iy={q\over 1-2i\e q}, \qquad dz = {dq\over \bar s(q)^2}
\end{equation}
The inner product integral now reads,
\begin{equation}
\delta_{nm}=\oint \phi_n(z) \phi_m(z) dz,
\end{equation}
where the integration path in the complex z-plane begins at $z=i/2\e$ 
($q=-\infty$) passes through the origin and once again ends at $z=i/2\e$
($q=\infty$)as shown in the figure.\cite{13}
\begin{figure}
\includegraphics[width=8.7cm]{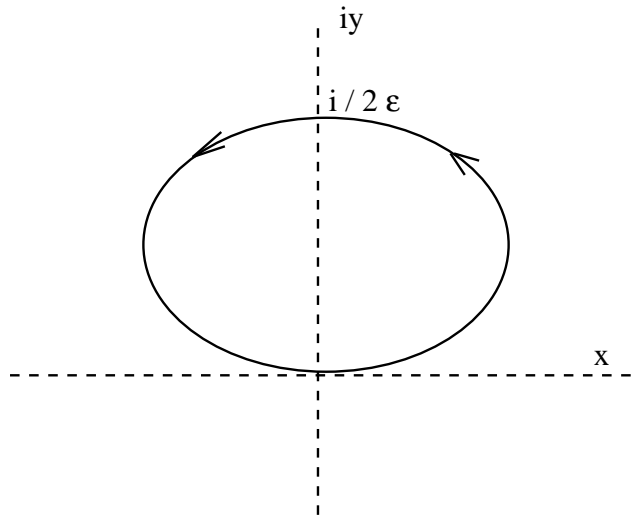}
\caption{The path required of the inner product of states Eq.(39) }
\label{path}
\end{figure}

\section{Summary}

We have found a non-trivial PT-symmetric continuum Hamiltonian, Eq.(9) which is 
isospectral to the simple harmonic oscillator with eigenvalues 
$n+1/2$. The corresponding eigenstates, Eq.(30), are PT-symmetric and 
form an orthonormal set of states, with path of integration a closed curve
in the complex plane. We have written down the action of the hidden symmetry
operator, $\cal C$, Eqs.(32,35) and have explicitly shown that its action on
any one of the states is to simply multiply it by its PT-norm; that is,
the ${\cal CPT}$ norm is real and postive. In contrast
to other systems studied in the continuum, we made no approximations.

\begin{acknowledgments}
It is a pleasure to thank Carl Bender for helpful discussions.  DG was supported by NSF grant PHY-2102914.
\end{acknowledgments}

\subsection*{Data availability statement}

No new data were created or analyzed in this study.


\begin{thebibliography}{99}
\bibitem{13} The length of this path is a finite number, $\pi/2\e$, which
only becomes infinite in the Hermitian limit $\e \to 0$.
\end{thebibliography}
\end{document}